\documentclass[final,5p,times]{amsci}

\usepackage{amssymb,amsfonts,amsmath,amstext,amsgen,amsopn,amsxtra,indentfirst,lscape}

\journal{Volume 100, Pages 334--341 (July/August 2012 Issue)}

\begin{document}

\begin{frontmatter}

\title{The Study of Climate on Alien Worlds}

\author{Kevin Heng}
\ead[eth]{heng@ias.edu}
\address{ETH Z\"{u}rich, Institute for Astronomy \\ Wolfgang-Pauli-Strasse 27, CH-8093, Z\"{u}rich, Switzerland}

\begin{abstract}
\textit{Characterizing atmospheres beyond our Solar System is now within our reach.\vspace{0.05in}\\
{\scriptsize Kevin Heng received his education in astrophysics (M.S. and Ph.D) at JILA (the Joint Institute for Laboratory Astrophysics) and the University of Colorado at Boulder. Subsequently, he was a postdoctoral researcher at the Institute for Advanced Study, Princeton, from 2007 to 2010 (including holding the Frank \& Peggy Taplin Membership from 2009 to 2010). He is currently a Zwicky Prize Fellow at the Institute for Astronomy at ETH Z\"{u}rich (the Swiss Federal Institute of Technology) in the Star and Planet Formation Group, where he is involved in the Exoplanet Characterization Observatory (EChO) mission proposed to the European Space Agency. He has worked on several topics in astrophysics, including shocks, planetesimal disks and fluid dynamics. His current, main research interest is in developing a hierarchy of theoretical tools to understand the basic physics and chemistry of exoplanetary atmospheres from the perspective of an astrophysicist. He spends a fair amount of time humbly learning the lessons gleaned from studying the Earth and Solar System planets, as related to him by atmospheric, climate and planetary scientists. He received a Sigma Xi Grant-in-Aid of Research in 2006.} \vspace{0.05in}\\
Text-only version of article, edited by Fenella Saunders.  Full version is available at: \texttt{www.americanscientist.org }
} 
\end{abstract}

\end{frontmatter}

It is a distracting, inconvenient coincidence that we are living in times of paradigm-shifting astronomical discoveries overshadowed by the deepest financial crisis since the Great Depression.  Amid a battery of budget cuts, the astronomical community has discovered more planets outside of our Solar System---called extrasolar planets or simply exoplanets---in the past decade than in previous millennia.  In the last couple of years alone, the Kepler Space Telescope has located more than 2,000 exoplanet candidates, including Earth-sized ones potentially capable of sustaining liquid water, demonstrating the ease at which nature seems to form them and hinting that we may be uncovering the tip of an iceberg.  Discovering and characterizing distant, alien worlds is an endeavor no longer confined to the realm of science fiction.  

In tandem with numerous surveys of the night sky performed from the ground, the Hubble, Kepler and Spitzer Space Telescopes observe the universe from outside of Earth's atmosphere.  These devices detect an exoplanet by recording the diminution of light as the body, residing in an edge-on orbit, passes in front of its host star. In the past few years, astronomers also have achieved the remarkable feat of measuring the diminution of light as the exoplanet passes behind its star, known as the secondary eclipse.  In other words, astronomical techniques have advanced to the point where we can detect a star masking the light from its exoplanet, which is a demonstrably small effect---at most a few parts in a thousand in the infrared and much smaller in the optical range of wavelengths. During a secondary eclipse, the light from an exoplanetary system originates only from the star, and these data can be used to subtract out the starlight when the exoplanet is not eclipsed.  All that remains is the light of the exoplanet and its atmosphere (if it exists). Such a technique has enabled astronomers to make the first detections of the light directly emitted by an exoplanet, which typically appears at its brightest in the infrared.  

Measuring transits and eclipses at several different wavelengths allows one to construct a spectrum of the exoplanetary atmosphere, of which a spectral analysis yields its composition and elemental abundances. (A spectrum describes the range of colors of the photons emanating from the exoplanet, but it generally extends beyond what our eyes can see toward both shorter and longer wavelengths.) In some cases, astronomers were able to record the ebb and rise of the brightness of the exoplanet as it orbits its parent star, otherwise known as the phase curve. An inversion technique, developed by Nick Cowan of Northwestern University and Eric Agol of the University of Washington, allows one to convert the phase curve into a ``brightness map," which is the latitudinally averaged brightness of the exoplanet across longitude.  Recent work by the same researchers has yielded two-dimensional information on the brightness of the exoplanet HD 189733b as a function of both latitude and longitude.  In other words, we have started to
do cartography on exoplanets!

\vspace{0.05in}
\noindent
\textbf{Tidal Locks}
\vspace{0.05in}

The first studies of exoplanetary atmospheres were performed on a class of objects known as hot Jupiters. A combination
of the transit technique with a measurement of the radial velocity (which is the gravitational wobble of a star as its exoplanet orbits around their common center of mass) yields the radius and mass of a hot Jupiter, respectively, and reveals that they are similar in these aspects to our own Jupiter.  The startling difference is that hot Jupiters are found about a hundred times
closer to their parent stars than Jupiter, which raises their surface temperatures to between 1,000 and 3,000 degrees
Kelvin. With spatial separations of a hundredth to a tenth of an astronomical unit (the average distance from the
Earth to the Sun) from their stars, the discovery of hot Jupiters caught the astronomical community by surprise, because their existence was neither predicted from astrophysical theory nor subsequently explained by it.  

Their large sizes render hot Jupiters easier to observe and thus the most obvious laboratories for extrasolar atmospheric studies.  Furthermore, the belief that their atmospheres are dominated by molecular hydrogen---which is consistent with the densities of the exoplanets, inferred from the astronomical observations to be about 1 gram per cubic centimeter---offers some hope that the atmospheres are primary, reflecting the composition of the primordial nebulae from which they formed, rather than secondary and reprocessed by geological mechanisms (such as on Earth).

Given enough time, an exoplanet's position and rotation tend to relax toward a state of minimum energy---a spin synchronized state, such that one hemisphere of the exoplanet always faces its parent star with the other hemisphere shrouded in perpetual darkness.  The characteristic time scale associated with this process is typically 1,000 times less than the age of the star.  (As a more familiar example, the Moon is in a spin synchronized state with respect to the Earth, notwithstanding its tiny rotational corrections called librations.)  In other words, one hot Jovian day is equal to one hot Jovian year. The unfamiliar configuration of permanent day- and night-side hemispheres on hot Jupiters opens up an unexplored regime of atmospheric circulation with no precedent in the Solar System and motivates theoreticians to test their tools in unfamiliar territory.  

Understanding these hot Jovian atmospheres requires clarifying the complex interplay between irradiation, atmospheric dynamics, chemistry and possibly magnetic fields. On the most irradiated hot Jupiters, the exoplanet viewed from the poles resembles a sphere painted half white and half black---the phase curve is a sinusoidal function that peaks at secondary eclipse and becomes dimmest at transit. Any shift of this peak from its reference point at secondary eclipse may be interpreted as being due to the presence of horizontal winds in the atmosphere, which act to transport heat from the day- to the night-side hemisphere. This angular shift was first measured for an exoplanet, the hot Jupiter HD 189733b, by Heather Knutson of the California Institute of Technology and her collaborators, who reported a peak shift of about 30 degrees east---in the direction of rotation. This angular shift was also measured for the hot Jupiters Ups And b (by Ian Crossfield of the University of California at Los Angeles and his collaborators) and WASP-12b (by Cowan and his collaborators).  

Other astronomers continue to push the envelope. Ignas Snellen of Leiden University and his colleagues, using the ground-based European Very Large Telescope (VLT), used a technique called absorption spectroscopy to measure the speed of the horizontal winds on the hot Jupiter HD 209458b. The technique compares the relative size of the exoplanet across a range of wavelengths.  At a wavelength where an atmospheric atom or molecule is the most absorbent, the exoplanet appears larger. By monitoring the shift in wavelength of an absorption line of carbon monoxide, the group determined that HD 209458b's winds clock in at about 2 kilometers per second, roughly 100 times faster than those on Earth. More attempts to measure atmospheric wind speeds are in the works, and these measurements remain at the cutting---if not the bleeding---edge of what astronomers can achieve.

\vspace{0.05in}
\noindent
\textbf{New Languages}
\vspace{0.05in}

The importance of these discoveries to astronomy cannot be overstated---they signal the dawn of exoplanetary meteorology, or at least legitimize its study in the eyes of astronomers and astrophysicists.  

Astronomers now possess a tool kit not only to measure the masses and sizes of exoplanets but also to characterize their atmospheric dynamics and chemistry.  Besides galvanizing the astronomical community, this newfound field is starting to exert a profound sociological impact on related fields of study: atmospheric and climate science, geophysics and planetary science. It marks the first great confluence of these fields with astrophysics, a gathering of scientists with different scientific and modeling philosophies, which is especially evident at interdisciplinary conferences where we struggle to understand one another's jargon. Atmospheric and climate scientists, as well as geophysicists, are firmly grounded in a data-rich regime, living within the system they study. Awash in an abundance of data from the terrestrial atmosphere and the geological record, no single model is capable of accounting for all of the observed phenomena. Instead, a hierarchy of models with different degrees of sophistication is utilized, with each model isolating key pieces of physics. The strategy is to first divide and conquer, then to unify and rule.

The knowledge gleaned from studying Earth and the Solar System planets serves as an invaluable guide, but there
is a cautionary tale to be told. As a rule of thumb, there are two characteristic length scales describing an atmosphere:
the Rhines length is the typical width of zonal (east-west) jets, whereas the Rossby length is the typical size of vortices
or eddies. For Solar System objects, both length scales are much smaller than planetary radii. On close-in exoplanets,
the Rhines and Rossby lengths are comparable to exoplanetary radii, implying that the atmospheric features are global in extent, an expectation that is borne out in three-dimensional simulations. The atmospheres of close-in exoplanets are thus in a circulation regime that is unprecedented in the Solar System.  Atmospheric circulation simulations therefore have to be global instead of local, and other physical implications---such as the mixing of atmospheric constituents and its effect on the
spectral appearance of the exoplanet---remain to be fully understood.  

The study of exoplanets is essentially confined to the scrutiny of point sources in the night sky.  Although we may obtain detailed spectral and temporal information on these point sources, the procurement of detailed spatial information remains a grand challenge for posterity.  Planetary scientists benefit from the ability to obtain photographs of the Martian surface and Jovian weather patterns, a privilege unavailable to astrophysicists.  It is important to recognize that astrophysicists are therefore trapped in a data-poor regime, with its myriad restrictions on how to construct models and interpret data. When faced with multiple explanations that are consistent with a given data set, astrophysicists often apply the principle of Occam's Razor: In the absence of more and better data, the simplest explanation is taken as the best one. To put it more tongue-in-cheek, one aims to be roughly accurate rather than precisely wrong. The need to recalibrate our scientific expectations and philosophies lies at the heart of this confluence of expertise.  

From studying the atmospheres of Earth and the Solar System planets, researchers have realized that atmospheres
are complex entities subjected to positive and negative feedback loops, exhibiting chemical, dynamical and radiative signatures over a broad range of time scales. Isaac Held of the Geophysical Fluid Dynamics Laboratory in Princeton, New Jersey, has argued that to truly understand these complex systems, one has to construct a hierarchy of theoretical models.  These simulations range from one-dimensional, pen-and-paper models that isolate a key piece of physics to full-blown, three-dimensional general circulation models (GCMs)---used for climate and weather forecasting---that incorporate a complicated soup of ingredients to capture the intricate interactions between the atmosphere, land and oceans on Earth.  For instance, GCMs concurrently solve a set of equations (called the Navier-Stokes equation) that treat the atmosphere as
a fluid, along with a thermodynamic equation and diverse influencing factors such as orography (mountain formation)
and biology. Many of these intricacies are unwarranted in theoretical investigations of exoplanetary atmospheres, and thus one of the key challenges is to realize how and when to simplify an Earth-centric model.  

Whether knowingly or unknowingly, a hierarchy of one- to three-dimensional theoretical models has emerged in the astrophysical literature.  Because the treatment of hot Jupiters and brown dwarfs---substellar objects not massive enough to sustain full-blown nuclear fusion at their cores---share several similarities, many of the pioneering models (by researchers such as Adam Burrows of Princeton University and Ivan Hubeny of the University of Arizona) were carried over from the latter to the former class of objects. Furthermore, the early models focused on the spectral appearance of hot
Jupiters, with the most sophisticated variants borrowing from an established technique in atmospheric and
climate science known as abundance and temperature retrieval.  Given the spectrum of an exoplanet, this technique
obtains the atmospheric chemistry and temperature-pressure profile consistent with the data. In the case of the hot Jupiter WASP-12b, Nikku Madhusudhan of Yale University and his collaborators inferred, using the retrieval technique, that the exoplanet possesses a carbon-to-oxygen ratio at least twice that of its star. If this result is confirmed---and the carbon-to-
oxygen ratio is measured for other exoplanets---it offers a valuable link between the properties of an exoplanetary
atmosphere and the formation history of the exoplanet.  

Astrophysicists have been quick to realize that atmospheric chemistry and dynamics intertwine in a nontrivial manner to produce the observed characteristics of a hot Jupiter. Adam Showman, a planetary scientist at the University of Arizona, became one of the first researchers to harness the power of GCMs in studying hot Jovian atmospheres. Several other researchers from the astrophysical community (including myself) followed soon after.  My collaborators and I generalized a benchmark test, which solves for the basic climatology of a (exo)planet using two methods of solution, to hot Jupiters. The transport of heat from the day-side to the night-side hemisphere of a spin synchronized exoplanet is---by definition---at least a two-dimensional problem. For extrasolar gas giants, the characteristic time scale on which the atmosphere reacts to such radiative disturbances spans many orders of magnitude, thus necessitating its theoretical consideration in three dimensions, an endeavor that is only tractable using GCMs.  Several groups have now successfully adapted GCMs to model exoplanetary atmospheres and are obtaining consistent results. Some outstanding technical issues remain, but it is clear that three-dimensional models are necessary if one wishes to predict not just the spectral appearance of exoplanets, but simultaneously their phase curves and temporal behavior. As the astronomical state-of-the-art advances, the exoplanets
being discovered will be more Earthlike, both in size and temperature, with the implication that GCMs will become even more relevant.

\vspace{0.05in}
\noindent
\textbf{Earthlike Exoplanets}
\vspace{0.05in}

We are only starting to understand the basic properties of hot Jupiters, including why some appear more inflated
than others and why some appear to redistribute heat more efficiently from their day-side to their night-side hemispheres.
In the cases of HD 189733b and HD 209458b, simulated spectra and phase curves---the latter of which constrains the efficiency of heat redistribution---that are computed using GCMs are able to match their observed counterparts fairly well. Until the next generation of space telescopes becomes operational, these examples remain the cornerstones of our understanding of exoplanetary atmospheres.  

HD 189733b is noninflated, meaning that its radius and mass are well matched by standard evolutionary theories of exoplanets (which predict the size of an exoplanet as it cools down from the primordial heat of formation).  It also appears to be shrouded in haze of unidentified chemistry, because its spectrum in the optical---obtained via the Hubble Space Telescope by Fr\'{e}d\'{e}ric Pont and David Sing of Exeter University, together with their collaborators---reveals a smooth, featureless slope consistent with Rayleigh scattering (the same process that causes the color of the sky as observed from Earth by preferentially affecting bluer sunlight).  

By contrast, HD 209458b is free of haze, but it is markedly larger than expected from evolutionary calculations. Theoretical ideas for radius inflation include the suggestion that partially ionized, hot Jovian atmospheres behave like giant electrical circuits, which when advected past an ambient magnetic field invoke Lenz's law on a global scale: Nature abhors a change in magnetic flux. Electric currents and opposing forces are induced to counteract the horizontal winds; the consequent conversion of mechanical energy into heat, called Ohmic dissipation, is believed to be responsible for keeping some hot Jupiters inflated. However, it remains to be proven if hot Jupiters even possess magnetic fields like those of Earth and
some of the Solar System planets. This field of research remains active. 

The study of hot Jupiters remains relevant because we already have the data to inform our hypotheses and modeling efforts, thereby affording us the opportunity to sharpen our theoretical tools---as much of the salient physics is identical---before applying them to Neptunelike or even Earthlike exoplanets, for which the data are currently scarce or nonexistent.  For many researchers, the ultimate prize is more familiar: to detect the spectrum of an Earthlike exoplanet orbiting a Sunlike star, and thereby answer age-old questions about the existence of extraterrestrial life. More succinctly, one wishes
to establish if the solitary example of an Earth twin in orbit around a solar twin is the only possible cradle for life
in the Universe. At the moment, such a quest remains elusive and appears out of the reach of even the next generation
of space telescopes.  

Instead, astronomers such as David Charbonneau of Harvard University and Jill Tarter of the SETI Institute have argued that a promising route toward detecting potentially habitable super Earths---Earthlike exoplanets with masses and radii somewhat larger than those of Earth---is to hunt for them around M stars (also known as red dwarfs).  These diminutive cousins of our Sun, with only a tenth to half of its mass, comprise about three-quarters of the stellar population in our galactic neighborhood.  There are several advantages to scrutinizing M stars: They are cooler in temperature than Sunlike stars, implying that their exoplanets may reside 10 to 100 times closer and yet still be able to harbor liquid water on their surfaces.  Being more proximate to their M stars renders these exoplanets more amenable to detection via current, established astronomical techniques---namely, transit and radial velocity measurements.  However, the price to pay is that they are expected to be spin synchronized and possess permanent day- and night-side hemispheres, much like their hot Jovian brethren. Such an expectation has led to theoretical concerns that their atmospheres may collapse due to the main constituent molecules condensing out on the frigid night sides. The astronomical approach to this conundrum is to charge forward with making new and better observations---after all, the answer is ultimately revealed by the data.  For example, for the super Earth GJ 1214b, transmission spectra have already been obtained, but interpretations about its atmospheric composition remain controversial.

\vspace{0.05in}
\noindent
\textbf{Better Telescopes}
\vspace{0.05in}

Experimentally, the next leap is to build dedicated, space-based telescopes capable of measuring high-resolution
spectra of exoplanets over protracted periods of time. Astronomers around the world are mobilizing to launch such
missions as the Exoplanet Characterization Observatory (EChO) and the Fast Infrared Exoplanet Spectroscopy Survey
Explorer (FINESSE), as proposed to the European Space Agency (ESA) and the National Aeronautics and Space Agency
(NASA), respectively. If and when these missions---or their successors---eventually fly (in the next decade or two), they
will deliver a bounty of both spectral and temporal information on hundreds of exoplanets, from which we may infer
their atmospheric chemistry, dynamics and climates. With a richly sampled data set of the emitted light from point-source
exoplanets over time, one may construct a power spectrum that elucidates the characteristic time scales on which
an exoplanet is flickering, indicating changes in temperature. Such a power spectrum of the atmosphere has been
spectacularly constructed for the Earth's surface, spanning time scales of under a day (diurnal variations) to many millennia
(called Milankovitch cycles, and inferred from the geological record).  Certainly, space missions are saddled with demands that will not allow for the construction of power spectra on time scales longer than a few months, but it is likely that many of the characteristic peaks in the power spectra will be compressed into a shorter time span for close-in exoplanets such as hot Jupiters and super Earths.  

The onus is on the theoretical community to lay down the foundation for understanding the climates of point-source exoplanets in general, thus moving us a step closer toward making more robust statements about their habitability.

\label{lastpage}


\begin{thebibliography}{99}

\bibitem[Burrows et al.(1997)]{burrows97} Burrows, A., et al. \ 1997, ``A Nongray Theory of Extrasolar Giant Planets and Brown Dwarfs", \textit{Astrophysical Journal}, 491, 856

\bibitem[Charbonneau et al.(2009)]{char09b} Charbonneau, D., et al. \ 2009, ``A super-Earth transiting a nearby low-mass star", \textit{Nature}, 462, 891

\bibitem[Fortney et al.(2010)]{fortney10} Fortney, J.J., Shabram, M., Showman, A.P., Lian, Y., Freedman, R.S., Marley, M.S., \& Lewis, N.K. \ 2010, ``Transmission Spectra of Three-dimensional Hot Jupiter Model Atmospheres", \textit{Astrophysical Journal}, 709, 1396

\bibitem[Guillot \& Showman(2002)]{gs02} Guillot, T., \& Showman, A.P. \ 2002, ``Evolution of 51 Pegasus b-like planets", \textit{Astronomy \& Astrophysics}, 385, 156

\bibitem[Held(2005)]{held05} Held, I.M. \ 2005, ``The Gap between Simulation and Understanding in Climate Modeling", \textit{Bulletin of the American Meteorological Society}, 86, 1609

\bibitem[Heng, Menou \& Phillipps(2011)]{hmp11} Heng, K., Menou, K., \& Phillipps, P.J. \ 2011, ``Atmospheric circulation of tidally locked exoplanets: a suite of benchmark tests for dynamical solvers", \textit{Monthly Notices of the Royal Astronomical Society}, 413, 2380

\bibitem[Heng, Frierson \& Phillipps(2011)]{hfp11} Heng, K., Frierson, D.M.W., \& Phillipps, P.J. \ 2011, ``Atmospheric circulation of tidally locked exoplanets: II. Dual-band radiative transfer and convective adjustment", \textit{Monthly Notices of the Royal Astronomical Society}, 418, 2669

\bibitem[Heng et al.(2012)]{hhps12} Heng, K., Hayek, W., Pont, F., \& Sing, D.K. \ 2012, ``On the effects of clouds and hazes in the atmospheres of hot Jupiters: semi-analytical temperature-pressure profiles", \textit{Monthly Notices of the Royal Astronomical Society}, 420, 20

\bibitem[Knutson et al.(2007)]{knutson07} Knutson, H.A., et al. \ 2007, ``A map of the day-night contrast of the extrasolar planet HD 189733b", \textit{Nature}, 447, 183

\bibitem[Madhusudhan et al.(2011)]{madhu11} Madhusudhan, N., et al. \ 2011, ``A high C/O ratio and weak thermal inversion in the atmosphere of exoplanet WASP-12b", \textit{Nature}, 469, 64

\bibitem[Peix\'{o}to \& Oort(1984)]{po84} Peix\'{o}to, J.P., \& Oort, A.H. \ 1984, ``Physics of climate", \textit{Reviews of Modern Physics}, 56, 365

\bibitem[Pierrehumbert(2010)]{p10} Pierrehumbert, R.T. \ 2010, ``Principles of Planetary Climate" (New York: Cambridge University Press)

\bibitem[Snellen et al.(2010)]{snellen10} Snellen, I.A.G., de Kok, R.J., de Mooij, E.J.W., \& Albrecht, S. \ 2010, ``The orbital motion, absolute mass and high-altitude winds of exoplanet HD 209458b", \textit{Nature}, 465, 1049

\bibitem[Seager \& Sasselov(2000)]{ss00} Seager, S., \& Sasselov, D.D. \ 2000, ``Theoretical Transmission Spectra during Extrasolar Giant Planet Transits", \textit{Astrophysical Journal}, 537, 916

\bibitem[Seager(2010)]{seager10} Seager, S. \ 2010, ``Exoplanet Atmospheres" (New Jersey: Princeton University Press)

\bibitem[Showman \& Guillot(2002)]{sg02} Showman, A.P., \& Guillot, T. \ 2002, ``Atmospheric circulation and tides of ``51 Pegasus b-like" planets", \textit{Astronomy \& Astrophysics}, 385, 166

\bibitem[Showman et al.(2009)]{showman09} Showman, A.P., Fortney, J.J., Lian, Y., Marley, M.S., Freedman, R.S., Knutson, H.A., \& Charbonneau, D. \ 2009, ``Atmospheric Circulation of Hot Jupiters: Coupled Radiative-Dynamical General Circulation Model Simulations of HD 189733b and HD 209458b", \textit{Astrophysical Journal}, 699, 564

\bibitem[Showman \& Polvani(2011)]{sp11} Showman, A.P., \& Polvani, L.M. \ 2011, ``Equatorial Superrotation on Tidally Locked Exoplanets", \textit{Astrophysical Journal}, 738, 71

\bibitem[Sing et al.(2011)]{sing11} Sing, D.K., et al. \ 2011, ``Hubble Space Telescope transmission spectroscopy of the exoplanet HD 189733b: high-altitude atmospheric haze in the optical and near-ultraviolet with STIS", \textit{Monthly Notices of the Royal Astronomical Society}, 416, 1443

\bibitem[Tarter et al.(2007)]{tarter07} Tarter, J., et al. \ 2007, ``A Reappraisal of the Habitability of Planets Around M Dwarf Stars", \textit{Astrobiology}, 7, 30

\end{thebibliography}
\end{document}